\documentstyle[amssymb,epsf,graphics,a4,12pt]{article}

\begin{document}
\newcommand{\beq}{\begin{equation}}
\newcommand{\eeq}{\end{equation}}
\newcommand{\beqn}{\begin{eqnarray}}
\newcommand{\eeqn}{\end{eqnarray}}
\newcommand{\dpf}{\displaystyle\frac}
\newcommand{\no}{\nonumber}
\newcommand{\ep}{\epsilon}

\begin{center}
{\Large\bf Modification to the power spectrum in the brane world inflation driven by the bulk inflaton}
\end{center}

\vspace{1ex} 
\centerline{\large Bin Wang$^{a,}$\footnote[1]{e-mail:binwang@fudan.ac.cn},
\ Li-Hui Xue $^{a}$, \ Xinmin Zhang$^{b,}$\footnote[2]{e-mail:xmzhang@mail.ihep.ac.cn}, and W-Y. P. Hwang$^{c}$
}

\begin{center}
{$^{a}$ Department of Physics, Fudan University, 200433 Shanghai. \\[0pt]
$^{b}$  Institute of High Energy Physics, Chinese Academy of Sciences,\\ P. O. Box 918(4), 100039 Beijing.  \\[0pt]
$^{c}$ Department of Physics, National Taiwan University, 106 Taipei. 
}
\end{center}

\vspace{6ex}

\begin{abstract}
We compute the cosmological perturbations generated in the brane world inflation driven by the bulk inflaton. Different from the model that the inflation is a brane effect, we exhibit the modification of the power spectrum of scalar perturbations due to the existence of the fifth dimension. With the change of the initial vacuum, we investigate the dependence of the correction of the power spectrum on the choice of the vacuum. 
\end{abstract}
\vspace{6ex} \hspace*{0mm} PACS number(s): 98.80.Cq, 11.10.Kk, 04.50.+h 
\vfill\eject

The standard cosmological model, based upon general relativity with an inflationary era, can potentially be used to explore Planck scale physics. The quantum field fluctuations at very early epochs were sensitive to short distance physics, and this sensitivity might leave imprint on the CMB spectrum produced by inflation. This observation has recently generated a lot of excitement about the possibility of opening a window on transplanckian or stringy physics in CMB anisotropies [1-17]. There are two major approaches to investigate this problem. One approach evolves from an attempt to use specific models of transplanckian physics, for example including models involving noncommutative geometry [14], and investigate its influence on the inflationary fluctuation spectrum. Another point of view has been taken with the focus on the choice of vacuum [15-17]. It has been shown that this method essentially amounts to an investigation of the physics of the de Sitter invariant vacua introduced in [18,19] and is a simple tool to study the effect of transplanckian physics. Though consistency of the de Sitter invariant vacua is questioned [20-22], it is argued that the application of the de Sitter invariant vacua to inflation cannot be excluded [23]. Due to the lack of complete quantum theory of gravity, it would be fair to say that a firm prediction of the signatures of transplanckian physics is still hard to be figured out at this moment.

Another significant interest of the cosmological perturbations generated during inflation is that it may provide a testing ground for the existence of extra dimensions. The initial work was done within the framework of a five-dimensional world where our visible universe is a three-dimensional brane located at a given point in the fifth dimension [24]. Considering the simplest possibility that the inflation is driven by a scalar field living on the three-dimensional brane, the effects of the transdimensional physics on the spectrum of the primordial density perturbations produced during the epoch of inflation have been studied. Though the power spectrum of the scalar perturbations is unchanged, the tensor perturbations receive a correction due to the existence of the fifth dimension. It is demonstrated that the cosmological perturbations are a powerful probe of the physics of extra dimensions.

The main purpose of the present paper is to extend the set up in [24] by assuming that the brane world inflation is driven by the bulk inflaton. We will compute the cosmological perturbations generated during the inflation and compare to those of the conventional four-dimensional inflationary scenario. Our findings indicate that the power spectrum of scalar perturbations of the inflation caused by the bulk inflaton receives modifications. We will also apply this result to the ansatz of Danielsson [15] for selecting a boundary condition for the mode function. The comparison to other discussions on quantum fluctuations in brane world inflation will be addressed in the end of the paper.

We consider a framework consisting of a four-dimensional brane embedded in a five-dimensional bulk with a stabilized radius. The metric for a natural brane universe model in the cosmological context takes the form [25,26]
\begin{equation} 
ds^2=-n^2 dt^2 + a^2 dX^2 + dy^2,
\end{equation}
where $n^2=(Hl)^2 \sinh ^2(y/l), a^2=a_0^2 n^2$ and $a_0=exp(Ht)$. Here $H$ is the four-dimensional Hubble constant and $y$ denotes the extra dimension. The four-dimensional hypersurface is at $y=y_0$[25] and the location of the brane is related with $H$ as follows
\begin{equation} 
H=\dpf{1}{l\sinh(y_0/l)},
\end{equation}
where $l=\vert 6/\Lambda_5\vert ^{1/2}$.

Due to the dynamics of the bulk gravitational field, the dilaton like scalar field with an effective potential $V(\phi)$ appears in the bulk. We assume $V$ is positive and may vary very slowly in space and time.The sufficiently slowly varying of the bulk scalar field will lead to the standard slow-roll inflation. As the universe evolves, the potential $V$ decreases and finally becomes zero and the standard Friedmann universe is recovered in the low energy limit. For simplicity, we take the model in [26] and  assume the potential of the form
\begin{equation} 
V(\phi)=V_0+\dpf{1}{2}m^2\phi^2
\end{equation}
with $m^2<0$ and consider the situation when $\phi$ is sufficiently close to zero. In the lowest order, $\phi=0$. The cosmological constant $\Lambda_{5,eff}$ in the bulk five-dimensional AdS spacetime is 
\begin{equation} 
\Lambda_{5,eff}=\Lambda_5+k_5^2 V_0
\end{equation}
and $\vert \Lambda_5\vert>k_5^2 V_0$ to ensure $\Lambda_{5,eff}<0$. The $l$ is replaced by $l_{eff}$ in the form $l^2_{eff}=6/\vert \Lambda_{5,eff} \vert$. In the following we write $l_{eff}$ into $l$ for simplicity.

The equation for a scalar field in the bulk background is given by
\begin{equation} 
-\dpf{1}{n^2}(3H\dpf{\partial \phi}{\partial t}+\dpf{\partial ^2\phi}{\partial t^2})+\dpf{1}{a^2}\dpf{\partial ^2 \phi}{\partial X^2}+\dpf{1}{n}(4\dpf{dn}{dy}\dpf{\partial \phi}{\partial y})+\dpf{\partial ^2 \phi}{\partial y^2}=m^2 \phi .
\end{equation}

Suppose in the momentum space the wave function can be seperated into 
\begin{equation} 
\phi(y,t,X)=\dpf{1}{(2\pi)^{3/2}}\int dk \dpf{f_{k}(y)g_{k}(\eta)}{a_0(\eta)}e^{ikX},
\end{equation}
where $\eta=-(H a_0)^{-1}=-exp(-Ht)/H$, so when $t\rightarrow -\infty, \eta\rightarrow -\infty$ and when $t\rightarrow \infty, \eta\rightarrow 0$.

Now the equations for $f(y)$ and $g(\eta)$ become
\begin{equation} 
\dpf{d^2 f(y)}{dy^2}+\dpf{4}{l}\coth(y/l)\dpf{d}{dy}f(y)-[m^2+\dpf{\lambda^2}{l^2 H^2 \sinh^2(y/l)}]f(y)=0
\end{equation}
\begin{equation} 
\dpf{d^2 g(\eta)}{d\eta^2}+[k^2-\dpf{1}{\eta^2}(2+\dpf{\lambda^2}{H^2})]g(\eta)=0
\end{equation}
where $\lambda^2$ is the separation constant corresponds to the Kaluza-Klein mass which appears in the Kaluza-Klein compactification.

The general solution of (7) is given by 
\beqn   
f(y)&=&A(Hlm)^{\gamma}(z^2-1)^{\gamma/2}{\rm F }\left( \frac{\gamma-\nu+2}{2},\frac{\gamma+\nu+2}{2},\gamma+\frac{5}{2};1-z^2 \right)  \\
&+&B(Hlm)^{-\gamma-3}(z^2-1)^{-(\gamma+3)/2}{\rm F }\left( -\frac{\gamma+\nu+1}{2},\frac{-\gamma+\nu-1}{2},-\gamma-\frac{1}{2};1-z^2 \right) \no,
\eeqn
where $z=\cosh(y/l), \nu=\sqrt{4+m^2l^2}, \gamma=-3/2+\mu$ and $\mu=\sqrt{9/4+\lambda^2/H^2}$.

Employing the properties of the hypergeometric function and the definition of Legendre polynomial, (9) can be reduced to
\beq 
f(y)=A¡¯ \frac{{\rm P }_{\nu-1/2}^{-\gamma-3/2}(z)}{\sinh^{3/2}(y/l)}
+ B¡¯ \frac{{\rm Q }_{\nu-1/2}^{-\gamma-3/2}(z)}{\sinh^{3/2}(y/l)}.
\eeq
As $z\rightarrow 1 (y\rightarrow 0)$, the Legendre function ${\rm Q }_{\nu-1/2}^{-\gamma-3/2}(z)$ is singular, which requires $B¡¯=0$ to keep the regularity. The coefficient $A¡¯$ in (10) is to be determined by the normalization condition [30]
\beq  
2(Hl)^2 \int_{0}^{y_0}dy \, \sinh^2(y/l)f_\lambda f_{\lambda '}^{*}=\delta(\lambda-\lambda '),
\eeq
which leads 
\beq   
A¡¯=\left\{ 2Hl \int_{0}^{y_0}dy \, \left[ \frac{{\rm P }_{\nu-1/2}^{-\gamma-3/2}\left(\cosh \frac{y}{l}\right)}{\sinh^{1/2}\frac{y}{l}} \right]^2 \right\}^{-1/2}.
\eeq

For $m=0$ and $\lambda=0$ case, the solution is a constant
\beq   
f_0=A¡¯=\frac{1}{Hl \sqrt{l \sinh(y_0/l)\cosh(y_0/l)-y_0}},
\eeq
which reflects the property on the four-dimensions and receives no influence from the extra dimension.

For $m\neq 0$, the contribution due to the extra dimension will emerge. The solution of (7) reads 
\beq   
f(y)=\left\{ 2Hl \int_{0}^{y_0}dy \, \left[ \frac{{\rm P }_{\nu-1/2}^{-\gamma-3/2}\left(\cosh \frac{y}{l}\right)}{\sinh^{1/2}\frac{y}{l}} \right]^2 \right\}^{-1/2}\frac{{\rm P }_{\nu-1/2}^{-\gamma-3/2}(\cosh(y/l))}{\sinh^{3/2}(y/l)}.
\eeq
Using the $Z_2$ symmetry
\begin{equation}  
\partial_y f(y)_{y=y_0}=0,  
\end{equation}
we find that $\gamma$ is determined by
\beq  
(\nu+2)z_0 P^{-\gamma-3/2}_{\nu-1/2}(z_0)=(\nu+\gamma+2)P^{-\gamma-3/2}_{\nu+1/2}(z_0)
\eeq
where $z_0=\cosh(y_0/l)$ and $y_0=l \sinh^{-1} (1/Hl).$

Eq(16) needs to be solved numerically. Given the location of the brane, $\gamma$ is determined by the $Z_2$ symmetry for each given set of $H, l, m$, which means that not all massive modes satisfy the $Z_2$ symmetry. With the obtained $\gamma$, the behavior of $f(y)$ on the brane $y=y_0$ can be got.

We now turn to study the time-dependent function $g(\eta)$. The general solution of (8) is expressed in the Bessel functions as
\beq  
g(\eta)=D_1 \eta^{1/2} J_{-\sqrt{9+4\lambda^2/H^2}/2}(k\eta)+D_2\eta^{1/2}J_{\sqrt{9+4\lambda^2/H^2}/2}(k\eta).
\eeq
When $k\eta>>1$, the asymptotic form of the mode function becomes
\beq      
g=e^{iE/2}[C_+ e^{i(k\eta-\pi/4)}+C_- e^{-i(k\eta-\pi/4)}],
\eeq
where $E=\pi\sqrt{9/4+\lambda^2/H^2}, C_+=D_1+D_2 e^{-iE}$ and $C_-=D_1 e^{-iE}+D_2$.

\begin{figure}[tbh]
\begin{center}
\leavevmode
\begin{eqnarray}
\epsfxsize= 11truecm\rotatebox{0}{\epsfbox{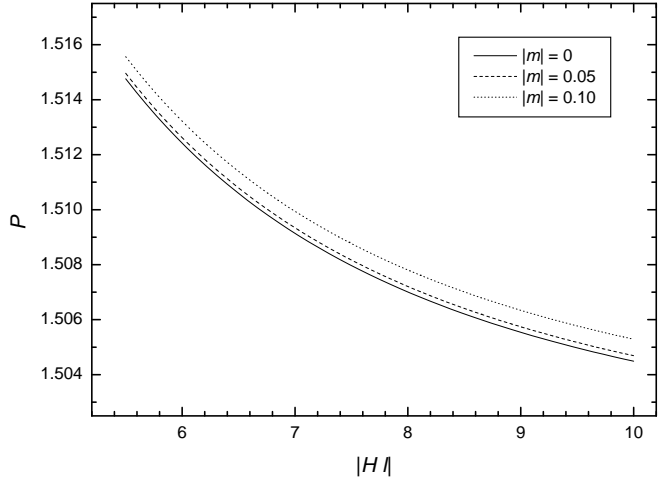}}  \nonumber
\end{eqnarray}
\vskip .5cm
\end{center}
\caption{\it{Modifications for scalar perturbations due to the extra dimension. The line for $m=0, \lambda=0$ is the spectrum generated in the conventional four-dimensional inflation. The other lines are for nonzero $m, \lambda$, which show spectrum in the brane world inflation. We take $H=1, l=1-10$ in the calculations.}}
\label{fig1}
\end{figure}
\begin{figure}[tbh]
\begin{center}
\leavevmode
\begin{eqnarray}
\epsfxsize= 11truecm\rotatebox{0}{\epsfbox{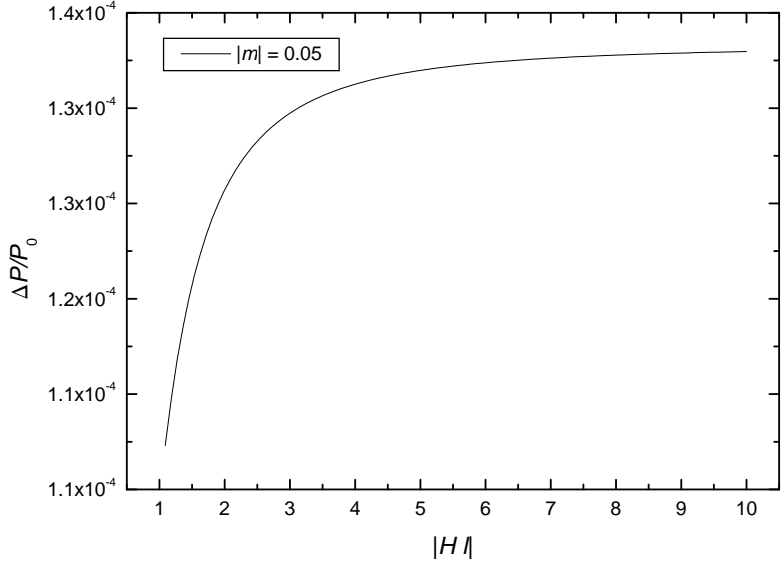}}  \nonumber
\end{eqnarray}
\vskip .5cm
\end{center}
\caption{\it{Modifications for scalar perturbations due to the extra dimension for $\vert m\vert=0.05$. }}
\label{fig2}
\end{figure}
\begin{figure}[tbh]
\begin{center}
\leavevmode
\begin{eqnarray}
\epsfxsize= 11truecm\rotatebox{0}{\epsfbox{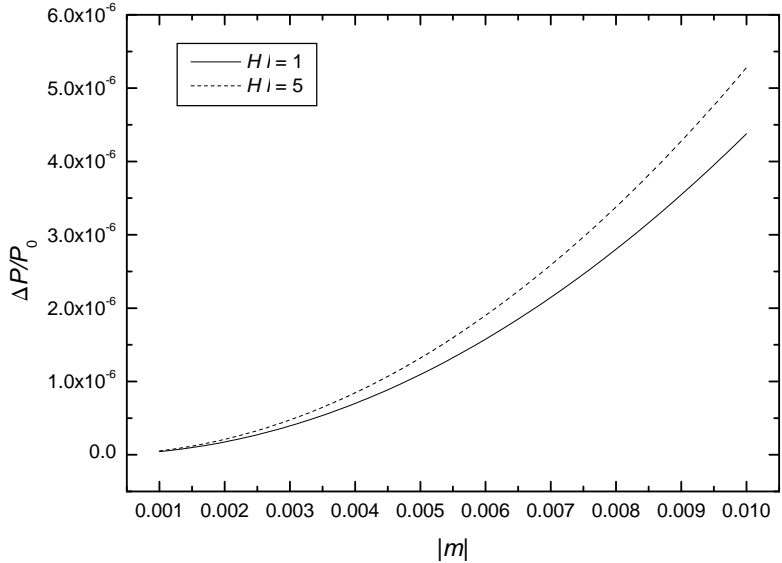}}  \nonumber
\end{eqnarray}
\vskip .5cm
\end{center}
\caption{\it{Dependence of the modification for the power spectrum due to extra dimension on values of $m$. }}
\label{fig3}
\end{figure}
\begin{figure}[tbh]
\begin{center}
\leavevmode
\begin{eqnarray}
\epsfxsize= 11truecm\rotatebox{0}{\epsfbox{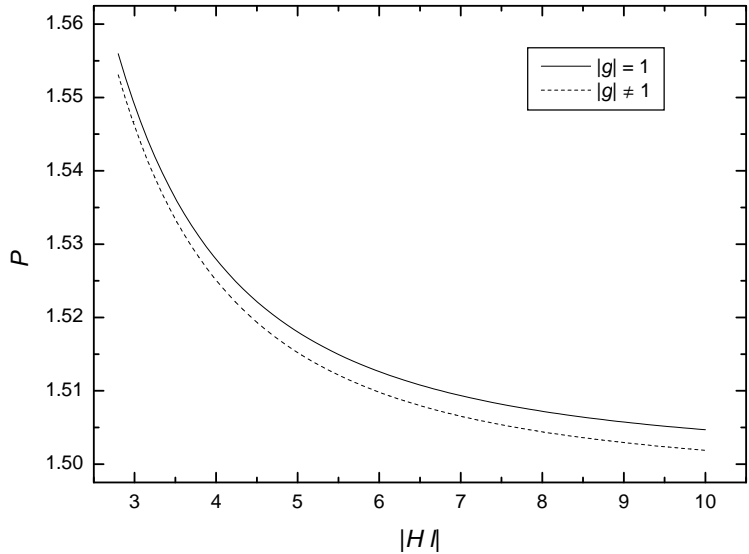}}  \nonumber
\end{eqnarray}
\vskip .5cm
\end{center}
\caption{\it{Modifications for scalar perturbations due to the choice of the vacuum. The black line is the spectrum by choosing the Bunch Davis vacuum and the dashed line is for de Sitter invariant $\alpha$ vacuum.}}
\label{fig4}
\end{figure}
\begin{figure}[tbh]
\begin{center}
\leavevmode
\begin{eqnarray}
\epsfxsize= 11truecm\rotatebox{0}{\epsfbox{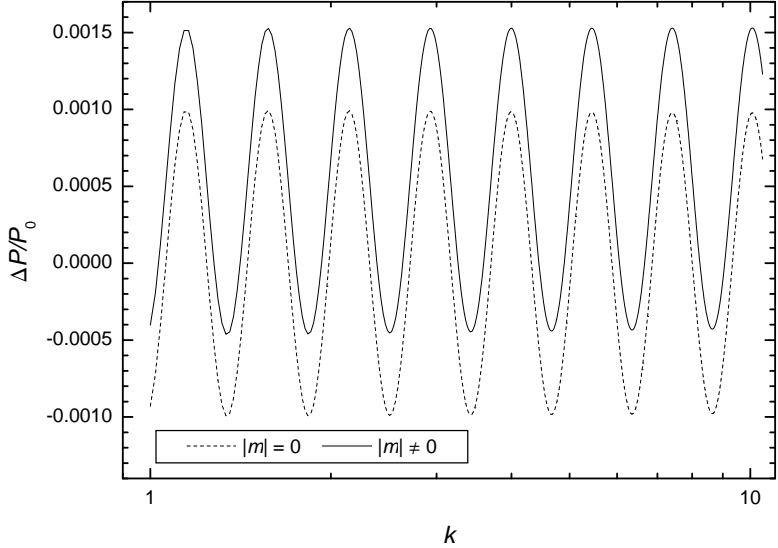}}  \nonumber
\end{eqnarray}
\vskip .5cm
\end{center}
\caption{\it{The modulation of the power spectrum in the transplanckian models for four dimensional inflation and brane world inflation.}}
\label{fig5}
\end{figure}
\begin{figure}[htb]
\begin{center}
\leavevmode
\begin{eqnarray}
\epsfxsize=8truecm\rotatebox{0}{\epsfbox{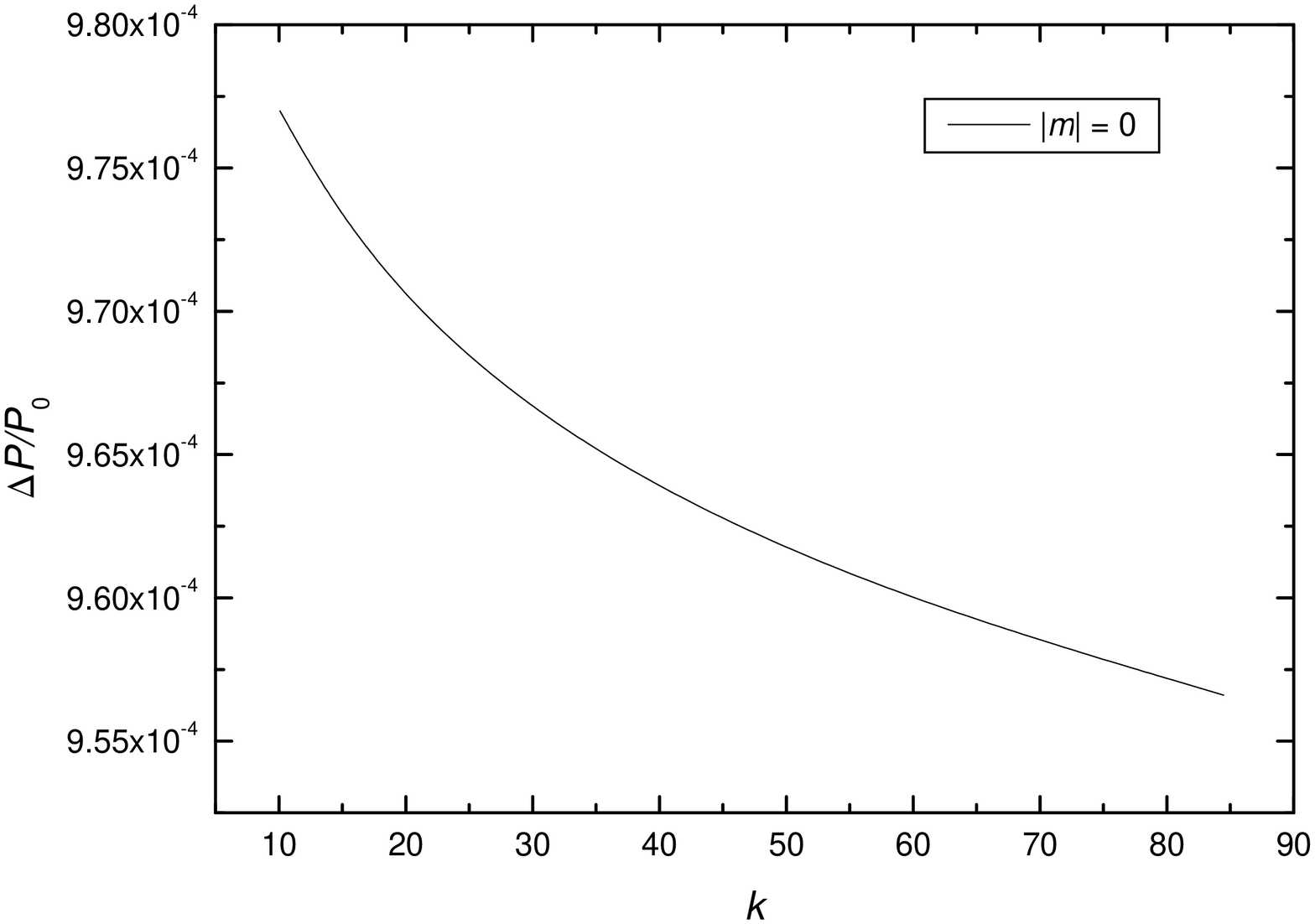}} & &
\epsfxsize=8truecm\rotatebox{0}{\epsfbox{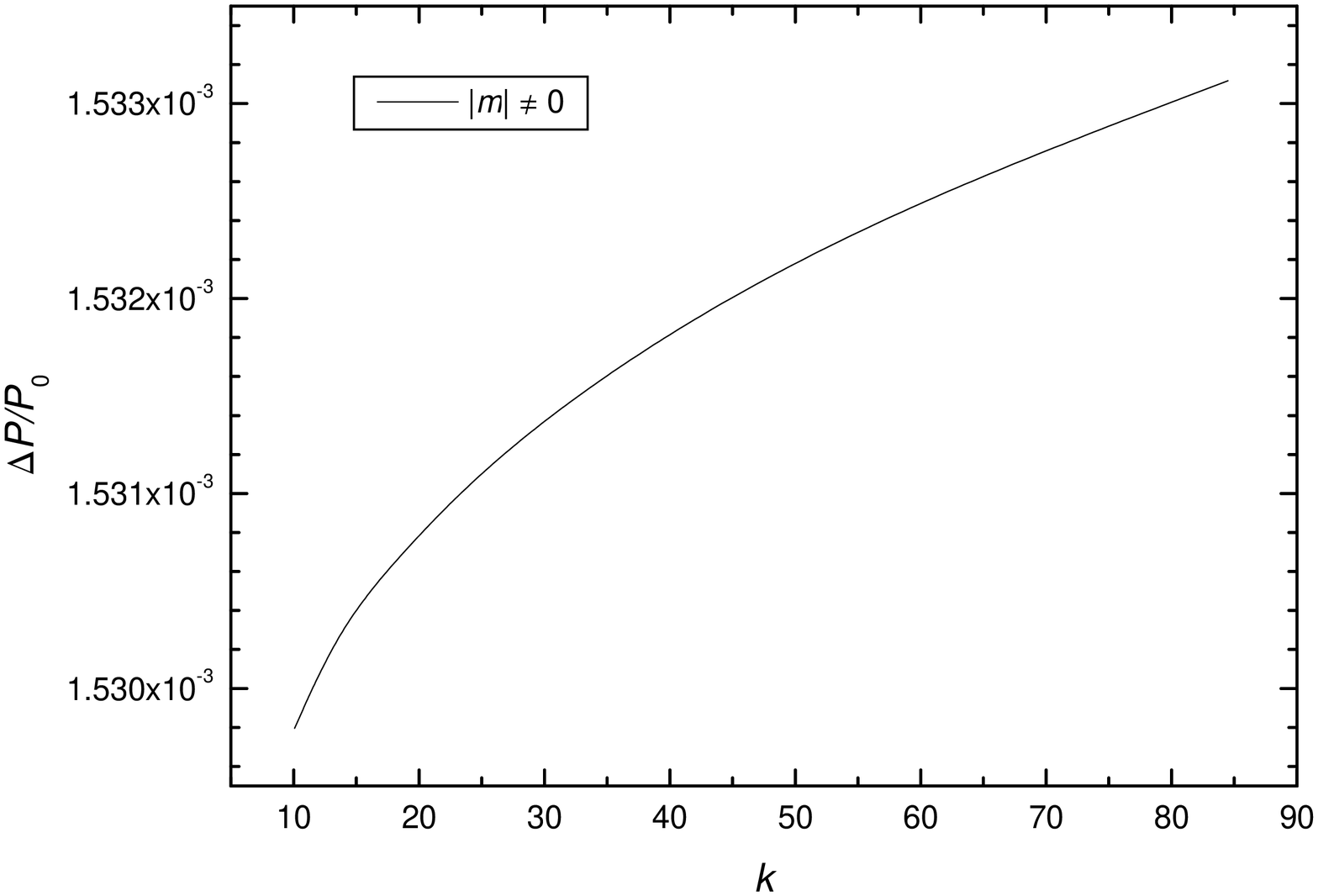}}\nonumber
\end{eqnarray}   
\vskip .5cm
\caption{\it{Lines connecting the points of wave crests in Fig.5 in four dimensional inflation and brane world inflation }}
\end{center}
\end{figure}

Danielsson has proposed a general ansatz for parametrizing the choice of vacuum in an inflationary spacetime [15]. The choice of Bunch-Davis vacuum can be expressed as a relation between the field and its conjugate momentum in the ultraviolat limit 
\beq    
\dpf{d\phi_p}{dt}\rightarrow -ip\phi_p, \hspace{1cm} p\rightarrow \infty.
\eeq
For an adiabatic vacuum, Danielsson enforced this relation at a finite momentum $p_c$,
\beq    
\dpf{d\phi_p}{dt}\rightarrow -ip\phi_p, \hspace{1cm} p=p_c.
\eeq
In the comoving variables, Danielssons condition is translated to
\beq    
\dpf{dg}{d\eta}=\dpf{g(i\eta k-1)}{\eta}.
\eeq

Considering (21) and the Wronskian condition $\vert C_+\vert^2-\vert C_- \vert^2=1$, we have 
\beq   
\vert C_+\vert^2=\dpf{1+4k^2\eta^2}{4k^2\eta^2}.
\eeq
We are free to choose the overall phase, so we can write the coefficients as 
\beqn  
C_+ & = & \dpf{2k\eta+i}{2k\eta}e^{-i(k\eta-\pi/4)} \no \\
C_- & = & \dpf{-i}{2k\eta}e^{i(k\eta-\pi/4)},
\eeqn
which agrees to the derivation in four-dimensions obtained in [15-17].

In the limit $k\eta\rightarrow \infty$, we have $C_+=1$ and $C_-=0$, which goes back to the Bunch-Davis vacuum.

The physical quantity of interest now is the contribution of the mode to the spectrum of CMBR perturbations. We compute this by examining $\vert\phi_4\vert^2$, where $\phi_4$ indicates the four-dimensional scalar field relating to the bulk inflaton $\phi$  by  $\phi_4=l^{1/2}\phi$,
\beqn      
<\phi_4^2> & = & l<\phi^2>_{y=y_0} \no \\
           & = & l\int d^3 k (\vert f\vert^2\vert g\vert^2)  \no \\
           & = & \int d^3 k P(k)
\eeqn
Let¡¯s first choose the Bunch-Davis vacuum, so that $\vert g\vert^2=1$. For $m=0, \lambda=0$ case, since there is no contribution due to the extra dimension, the power spectrum of the physical expectation value should be that of the four dimensional case, i.e. $P(k)=\dpf{1}{(Hl)^2(l\sinh(y_0/l)\cosh(y_0/l)-y_0)}$. For nonzero $m, \lambda$, the present day power spectrum of scalar modes generated by a primordial period of inflation on our visible brane at $y=y_0$ can be calculated from (14) and (16). The results are shown in Fig.1. In Fig.2, we plot the ratio 
\beq   
\dpf{\Delta P}{P_0}=\dpf{P-P_0}{P_0},
\eeq
where $P_0$ and $P$ represent power spectrums for zero and nonzero $m$, respectively.
It is clear to see that the spectrum of scalar perturbations has been modified due to the existence of the fifth dimension.  The bigger value of $m$ is, the more change to the power spectrum we have. The dependence of the modification to the power spectrum on the value of $m$ is more explicitly shown in Fig.3. Comparing to the simple case that the inflation is only a brane effect [24], we have observed here that the modification of the power spectrum of scalar perturbations in the brane world inflation driven by the bulk inflaton. In the numerical calculation we have taken the slow-roll condition $\vert m^2\vert\ll H^2$ [26].

Adopting the de Sitter invariant $\alpha$-vacua as plausible initial conditions for inflation, from (18) and (23) we have 
\beq    
\vert g\vert^2 = 1-\dpf{H}{M}\sin(2M/H)
\eeq
where $M$ is the energy scale of the new physics, e.g. the Planck scale or the string scale, and $H$ is the Hubble constant during inflation. Calculating the spectrum of the CMBR perturbations, in Fig.4 we have shown that the perturbation spectrum for this new vacuum deviated from that of the Bunch-Davis vacuum. This effect appears both in the conventional four-dimensional inflation and the brane world inflation. 

As shown in (26) and Fig.4, a Hubble constant which does not change during inflation would just imply a small change in the overall amplitude of the power spectrum. This would not constitute a useful signal. It is expected that the situation will be more interesting if the Hubble constant is to vary. As argued in [31], using the slow roll parameter $\epsilon$ and evaluating $H$ when a given mode crossing the horizon($k=aH$), $H\sim k^{-\epsilon}$. The correction in (26) $H/M=\xi (k/k_n)^{-\epsilon}$, where $k_n$ is the particular scale leaves the horizon corresponding to the largest angular scales measurable in the CMBR and $\xi=H_n/M\sim 4\times 10^{-4}\sqrt{\epsilon}/\gamma$, where $\gamma=M/M_{Pl}$. The power spectrum in the new vacuum can be parametrized in the form
\beq  
P(\epsilon,\xi,k)=P_0(k)\left(1-\xi(\dpf{k}{k_n})^{-\epsilon}\sin[\dpf{2}{\xi}(\dpf{k}{k_n})^{\epsilon}]\right),
\eeq
where $P_0(k)$ is the scale invariant spectrum which we can use for comparison in the numerical work.

Adopting reasonable values of $\epsilon, \gamma$ as $\epsilon=0.01, \gamma=0.01$ [31], in Fig.5 we show examples of power spectrum for zero and nonzero values of $m$. The dotted line shows the modulation of the power spectrum predicted in the transplanckian model for four dimensional inflation with zero $m$. And the solid line shows the transplackian effect in the brane world inflation with nonzero $m$. Connecting the points of wave crests, we exhibit again in Fig.6 the difference between the four dimensional inflation and the brane world inflation. To connect to observable quantities from satellite like MAP and Planck, we have to compute the angular power spectrum $C_l$ for each set of parameters. This is a complicated task which has to be performed numerically. We will leave it for future studies. 

In summary we have investigated the cosmological perturbations generated in brane world inflation driven by the bulk inflaton. In the conventional four-dimensional inflationary scenario, the relation between the density perturbation and fluctuation of the scalar field is well known [27]. However the scalar field living in the bulk has both $t$ and $y$ dependence. The analysis of the cosmological fluctuation becomes much more involved [28] and there exists the non-trivial background solution due to the $y$ dependence. The density perturbation may change a lot compared with that of the four-dimensional case. We have shown that the perturbations are a powerful tool to explore the physics of extra dimensions. Different from the result of the inflation taking place on the brane, modifications of the power spectrum of scalar perturbations have been exhibited in the inflation driven by the bulk inflaton. This result consolidate the argument given in [24].

Besides assuming that the initial vacuum is the traditional Bunch-Davis vacuum, we have also studied the transplanckian physics by introducing the de Sitter invariant $\alpha$-vacuum. The correction scaling as $H/M$ as suggested by the analysis performed in the four-dimenisonal cases has also been observed in the brane world inflation.   

Finally we would like to compare our result to recent discussions of the quantum fluctuations in brane world inflation [29,30]. In [29], the quantum fluctuations of a massless bulk scalar field $\phi$ in the $AdS_5$ background was investigated. They decomposed the bulk field $\phi$ into the zero mode and the Kaluza-Klein (KK) modes with the zero mode existing on the brane while the KK mode being the contribution from the extra dimension. It was found in [29] that the KK contribution is too small relative to the zero mode contribution. This discussion was extended to a bulk scalar field of arbitrary mass [30] and the non-negligible KK mode compared to that of zero mode was observed. With nonzero mass of the bulk field, the extra dimensional influence appears. In [30] both zero mode and the KK mode contain extra dimensional contribution (see (3.14) in [30] for their zero mode). Thus they compared extra dimensional contributions due to different modes to the brane. However in our study we have examined the difference between the fluctuation spectrum in four-dimensions and the contribution from the extra dimension. We found that the deviation of the scalar perturbation spectrum from that of the four-dimensional case caused by the nonzero mass bulk scalar field  exists. This modification could be the probe of the existence of the extra dimension. Another difference between our study and the work in [29,30] is that we have also considered the modification of the power spectrum due to the change of the initial vacuum, while in their work only the fixed vacuum was taken.

ACKNOWLEDGEMENT: We are grateful to R. Brandenberger, R. J. Zhang, M. Z. Li, E. Abdalla and B. Cuadros-Melgar for valuable discussions. This work was partially supported by the Ministry of Science and Technology of China under grant No. NKBRSF G19990754, NNSFC grant 10005004 and the foundation of Department of Education of China. B. Wang would also like to acknowledge the NSF grant PHY99-07949 at the Kavli Institute for Theoretical Physics, UCSB, where the work was initiated.


\begin{thebibliography}{99}
\bibitem{} R. H. Brandenberger, hep-th/9910410; J. Martin and R. H. Brandenberger, Phys. Rev. D 63, 123501 (2001); R. H. Brandenberger and J. Martin, Mod. Phys. Lett. A 16, 999 (2001); J. Martin and R. H. Brandenberger, astro-ph/0012031; R. H. Brandenberger, S. E. Joras and J. Martin, hep-th/0112122; J. Martin and R. H. Brandenberger, hep-th/0201189; R. H. Brandenberger and J. Martin, hep-th/0202142.
\bibitem{} J. C. Niemeyer, Phys. Rev. D 63, 123502 (2001); J. C. Niemeyer and R. Parentani, Phys. Rev. D 64, 101301 (2001); A. Kempf and J. C. Niemeyer, Phys. Rev. D 64, 103501 (2001); J. C. Niemeyer, astro-ph/0201511; J. C. Niemeyer, R. Parentani and D. Campo, hep-th/0206149.
\bibitem{} A. A. Starobinsky, JETP Lett. 73, 371 (2001).
\bibitem{} A. Kempf, Phys. Rev. D 63, 083514 (2001).
\bibitem{} L. Mersini, M. Bastero-Gil and P. Kanti, Phys. Rev. D 64, 043508 (2001).
\bibitem{} M. Bastero-Gil and L. Mersini, Phys. Rev. D 65, 023502 (2002).
\bibitem{} L. Hui and W. H. Kinny, astro-ph/0109107.
\bibitem{} M. Bastero-Gil, P. H. Frampton and L. Mersini, hep-th/0110167.
\bibitem{} G. Shiu and I. Wasserman, hep-th/0203119.
\bibitem{} S. Shankaranarayanan, gr-qc/0203060.
\bibitem{} S. F. Hassan and M. S. Sloth, hep-th/0204110.
\bibitem{} N. Kaloper, M. Kleban, A. E. Lawrence and S. Shenker, hep-th/0201158.
\bibitem{} R. Easther, B. R. Greene, W. H. Kinny and G. Shiu, hep-th/0110226.
\bibitem{} R. Brandenberger and P. M. Ho, hep-th/0203119; F. Lizzi, G. Mangano, G. Miele and M. Peloso, hep-th/0203099.
\bibitem{} U. H. Danielsson, Phys. Rev. D 66, 023511 (2002).
\bibitem{} R. Easther, B. R. Greene, W. H. Kinny and G. Shiu, Phys. Rev. D 64, 103502 (2001); hep-th/0204129.
\bibitem{} U. H. Danielsson, JHEP 0207, 040 (2002); K. Goldstein and D. Lowe, hep-th/0208167.
\bibitem{} N. A. Chernikov and E. A. Tagirov, Ann. Inst. Henri Poincare, Vol. IX, 109 (1968); E. Mottola, Phys. Rev. D 31, 754 (1985); B. Allen, Phys. Rev. D 32, 3136 (1985); R. Floreanini, C. T. Hill and R. Jackiew, Annals Phys. 175, 345 (1987).
\bibitem{} R. Bousso, A. Maloney and A. Strominger, hep-th/0112218; M. Spradlin and A. Volovich, hep-th/0112223.
\bibitem{} T. Banks and L. Mannelli, hep-th/0209113.
\bibitem{} M. B. Einhorn and F. Larsen, hep-th/0209159.
\bibitem{} N. Kaloper, M. Kleban, A. Lawrence, S. Shenker and L. Susskind, hep-th/0209231.
\bibitem{} U. H. Danielsson, hep-th/0210058.
\bibitem{} G. F. Giudice, E. W. Kolb, J. Lesgourgues and A. Riotto, Phys. Rev. D 66, 083512 (2002).
\bibitem{} J. Garriga and M. Sasaki, Phys. Rev. D 62, 043523 (2000).
\bibitem{} Y. Himenmoto and M. Sasaki, Phys. Rev. D 65, 104020 (2002).
\bibitem{} R. Brandenberger and J. Martin, Int. J. Mod. Phys. A 17, 3663 (2002).
\bibitem{} C. van de Bruck, M. Dorca, R. Brandenberger and A. Lukas, Phys. Rev. D 62, 123515 (2000).
\bibitem{} S. Kobayashi, K. Koyama and J. Soda, Phys. Lett. B 501, 157 (2001).
\bibitem{} N. Sago, Y. Himemoto and M. Sasaki, Phys. Rev. D 65, 024014 (2002).
\bibitem{} L. Bergstrom and U. H. Danielsson, hep-th/0211006.
\end{thebibliography}
\end{document}